\documentclass[aip, pop, twocolumn, showpacs, letterpaper, 10pt]{revtex4-1}

\usepackage{array}
\newcolumntype{P}[1]{>{\centering\arraybackslash}p{#1}}
\newcolumntype{M}[1]{>{\centering\arraybackslash}m{#1}}
\usepackage{multirow}
\usepackage{color}
\usepackage{amsmath}
\usepackage{amssymb}
\usepackage{graphicx}
\usepackage{xspace}

\begin{document}
\newcommand{\Eq}[1]{Eq.~(\ref{#1})}
\newcommand{\Eqs}[2]{Eqs.~(\ref{#1},\ref{#2})}
\newcommand{\Fig}[1]{Fig.~(\ref{#1})}
\newcommand{\Sec}[1]{Sec.~(\ref{#1})}
\newcommand{\PD}[2]{\frac{\partial #1}{\partial #2}}
\newcommand{\Skn}[2]{\sum\limits_{#1=0}^{#2-1}}
\newcommand{\rvec}[1]{|#1\rangle}
\newcommand{\lvec}[1]{\langle#1 |}
\newcommand{\sprod}[2]{\langle#1|#2\rangle}

\title{Simulation of FuZE axisymmetric stability using gyrokinetic and extended-MHD models}

\author{V.~I.~Geyko, J.~R.~Angus, and M.~A.~Dorf}
\affiliation{Lawrence Livermore National Laboratory, Livermore, California, 94550, USA}

\begin{abstract}
Axisymmetric ($m=0$) gyrokinetic and extended-MHD simulations of sheared-flow Z-pinch plasma are performed with the high-order finite volume code COGENT. The present gyrokinetic model solves the long-wavelength limit of the gyrokinetic equation for both ion and electron species coupled to the electrostatic gyro-Poisson equation for the electrostatic potential. The electromagnetic MHD model includes the effects of the gyro-viscous pressure tensor, diamagnetic electron and ion heat fluxes, and generalized Ohm's law. A prominent feature of this work is that the radial profiles for the plasma density and temperature are taken from the FuZE experiment and the magnetic field profile is obtained as a solution of the MHD force balance equation. Such an approach allows to address realistic plasma parameters and provide insights into the current and planned experiments. In particular, it is demonstrated that the radial profiles play an important role in stabilization, as the embedded guiding center ($E{\times} B$) drift has a strong radial shear, which can contribute to the Z-pinch stabilization even in the absence of the fluid flow shear. The results of simulations for the FuZE plasma parameters show a decrease of the linear growth rate with an increase in the flow shear, however full stabilization in the linear regime is not observed even for large (comparable to the Alfv\'en velocity) radial variations of the axial flow. Nonlinear stability properties of the FuZE plasmas are also studied and it is found that profile broadening can have a pronounced stabilizing effect in the nonlinear regime. 

\end{abstract}

\date{\today}

\maketitle

\section{Introduction}
\label{sec:intro}

From the dawn of magnetic fusion energy studies, the Z-pinch concept has been considered as a possible plasma confinement configuration suitable for maintaining a controlled fusion reaction\cite{reynolds52,kurchatov57}. The Z-pinch configuration is a cylindrically symmetric plasma column with an axial current inside, such that the generated magnetic filed creates an inward Lorentz force that confines the plasma. Relatively simple cylindrical geometry and the absence of any external magnetic fields together with a great utilization of the generated magnetic field ($\beta\sim 1$) make this concept quite attractive. However, Z-pinch plasmas are susceptible to rapid  magnetohydrodynamics (MHD) instabilities, whose growth rate $\gamma_i$ is on the order of the inversed Alfv\'en time $\gamma_i \sim V_a/a$, where $V_a$ is the Alfv\'en speed and $a$ is the characteristic radial size of the pinch. The instabilities completely disrupt the pinch as was observed in certain experiments\cite{kurchatov57}. The local linear MHD analysis of these instabilities was done by Kadomtsev\cite{kadomtsev60}, and it was shown that the most unstable modes are $m=0$ and $m=1$, called \textit{sausage} and \textit{kink} modes respectively, where $m$ is the angular number. Apart from MHD modes, which typically have a spatial scale of the pinch radius, short wavelength drift modes with scale on the order of ion gyroradius $\rho_i$ and growth rate on the order\cite{ricci06pop,ricci06prl} $\gamma_d \sim (k_\perp \rho_i)V_a/a$ can develop as well. Here,  $k_\perp$ is the wave vector in the perpendicular to the magnetic field direction. These modes appear naturally in gyrokinetic formulation\cite{ricci06pop,simakov01} and can also be captured with extended-MHD models\cite{kadomtsev60,angus19} if proper drift terms are retained. For the wavelengths on the order of the ion gyroradius, these modes can be as destructive as the ideal MHD modes, therefore the problem of Z-pinch stabilization becomes even more complicated.

A renewed interest to the concept was prompted by the recent successful experiments on the sheared flow stabilized (SFS) Z-pinches, namely, ZaP\cite{shumlak01,shumlak03} and FuZE\cite{zhang19,shumlak20}. The most recent FuZE experiment reports a pinch with a current of $200$~kA stable for approximately 5000 Alfv\'en time scales, which is drastically greater than characteristic linear instability growth time. In both experiments, an axially sheared plasma flow is believed to play a key stabilizing role. While there are no reported values of shear for the FuZE, the ones from the ZaP experiment are somewhat a fraction of the Alfv\'en velocity over the pinch radius\cite{golingo05} $a$.

The idea of using a sheared flow for stabilization of Z-pinch plasmas originates from the early work of Shumlak and Hartman\cite{shumlak95}, where they demonstrated that a moderate shear is capable of stabilizing the $m=1$ MHD mode, provided $dv_z/dr\ge 0.1k V_a$, where $k$ is the axial wave vector and $v_z$ is the plasma flow velocity. This result was obtained for the Kadomtsev profile\cite{kadomtsev60}, which is marginally stable against the $m=0$ MHD mode. However, these results are in disagreement with Arber's work\cite{arber96} where no pronounced mitigation of the $m=1$ mode with the wavelength $ka=10/3$ was observed even for larger values of flow shear. A more detailed recent study\cite{angus20dg} on the stability of linear ideal MHD modes did not support the original hypotheses either.
 Nonlinear ideal-MHD simulations of the $m=0$ mode were done by Parischev\cite{paraschiv10}, and a different stabilization condition was reported $dv_z/dr\ge V_a/a$. The ideal MHD model is, however, of limited validity because of relevant experimental parameters for which (i) the plasma is hot and therefore not strongly collisional, (ii) the ion Larmor radius is not infinitesimally small, $\rho_i/a\sim 0.1$, and therefore, finite Larmor radius (FLR) effects have to be taken into account. To overcome these issues, extended-MHD\cite{angus19}, gyrokinetic\cite{geyko19}, and fully kinetic\cite{tummel19} simulations were performed by different authors. The main difference between the kinetic and ideal-MHD models is that the linear growth rate can decrease for $k\rho_i\gtrsim 1$ due to FLR effects\cite{arber94}, which is not observed in the ideal MHD simulations. The stabilizing effect by a sheared flow on the $m=0$ mode was observed in all the simulations, yet, no complete stabilization by the moderate shear was demonstrated in the gyrokinetic simulations\cite{geyko19}. Furthermore, the stabilization of the $ka=5.0$ mode only was reported in the fully kinetic simulations\cite{tummel19}, and no data for other wavelengths was obtained.

The aforementioned simulations were performed for the case of some special model profiles for density and temperature. The most common choice was the diffuse Bennett\cite{bennett34,bennett55} profile, which has a unique property of being an equilibrium solution for a fully kinetic formulation. A noticeable feature of this profile is that it has a moderate logarithmic derivative of the pressure, and depending on the adiabatic gas index $\Gamma$ the profile is either stable or unstable at all $r$ for the $m=0$ mode\cite{angus19}. A \textit{realistic} profile obtained from recent experimental data\cite{zhang19} is however drastically different from the Bennett model profile. Therefore the stability properties of the FuZE plasmas can be substantially different from those obtained in the previous numerical studies.

In the present paper, we make use of the COGENT\cite{dorf13} code to simulate the $m=0$ mode behavior in a realistic (also called \textit{FuZE-like}) type of pinch profiles. 
The simulations are performed by making use of the electrostatic and extended-MHD simulation models. The gyrokinetic formulation employs the electrostatic full-F long-wavelength approximation. This model was tested and compared\cite{geyko19} to fully kinetic PIC simulations, and it was shown to adequately capture physics related to FLR effects. The model is missing electromagnetic effects and higher order FLR effects, as well as the capability to deal with sonic-range flow velocities. The extended MHD model includes a gyroviscous pressure tensor based on Braginskii formulation, generalized Ohm's law and diamagnetic electron and ion heat fluxes in the energy density equations. As mentioned earlier, the MHD model has a limited validity for the parameters characteristic to the FuZE plasmas. Nevertheless, its formulation consistently include electromagnetic effects and allows for arbitrary large shear values. While a better extension of the model is needed in order to correctly capture collisionless ion FRL physics (for example, a CGL model can be considered\cite{chew56}), the results can still provide good insights on the plasma behavior and shear flow stabilization process. Both models demonstrate that a moderately sheared flow is not sufficient to provide a global pinch stabilization. In addition, the effects of a profile shape on the stability properties have been addressed and it is found that profile broadening can have a pronounced stabilizing effect in the nonlinear regime. 
 
The paper is organized as follows. \Sec{sec:theory} contains theoretical background including data fitting, scaling analysis, and a review on the main results obtained in the previous work. Details and results of the gyrokinetic simulations are shown in \Sec{sec:gk_sec}. The MHD model equations and simulations are covered in \Sec{sec:mhd_sec}. Speculations on possible stabilization mechanism for broad pinch profiles are provided in \Sec{sec:flatten}. In \Sec{sec:concl}, the main results are summarized.

\section{Theoretical background}
\label{sec:theory}
\subsection{Realistic profiles}

We define a set of functions that describe the radial dependence of the plasma density $n(r)$ and temperature $T(r)$ as a ``FuZE-like profile'' if these functions are obtained via a nonlinear curve fitting of the FuZE experimental data. The data is provided in the recent work of Zhang \textit{et al}.\cite{zhang19}. In particular, the density and temperature data are shown in Fig.~3b and Fig.~4b of the cited work respectively. Ignoring error bars, the data is fitted with smooth analytical functions in order to be used as the initial conditions for numerical simulations. 

\begin{figure}
	\centering
	\begin{center}
			\includegraphics[width=0.45\textwidth]{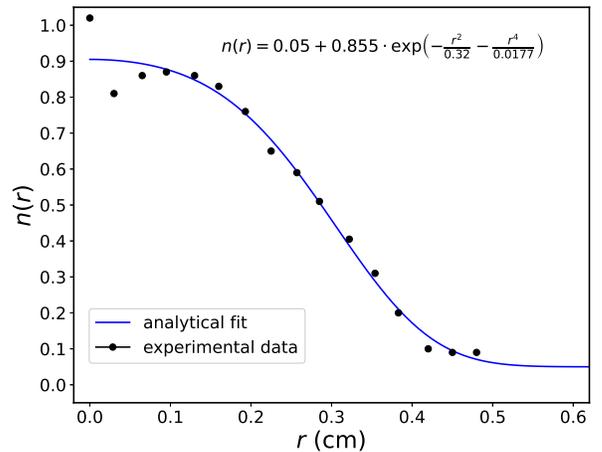}
	\end{center}
	\caption{Least square density fitting with a smooth function. Data points are taken from the experimental results\cite{zhang19}. Error bars are ignored and all points are taken with the same weights.}
	\label{fig:d_fit}
\end{figure}

While there are two profiles for density provided in Ref. \cite{zhang19}, measured at axial locations of $z=13.8$~cm and $z=15.0$~cm, only the former is used in the present work. The most noticeable difference between them is observed at the interior of the pinch close to the axis, yet the simulated mode of interest is localized close to the periphery, thus making this difference negligible. Moreover, a smooth fitting is not possible at $r=0$ even if the measurement errors are considered. Therefore using both profiles is unnecessary and the only one is picked for all the simulations here.

There is an infinite amount of possible fitting functions, yet, they have to satisfy the following constraints in order to obtain a reasonable setup. First, the radial derivative of the total pressure at $r=0$ should be zero, otherwise the curl of magnetic field has an irregular point, hence a divergent current density on the axis of the pinch (See \Sec{sec:flow} for more details). Second, since the simulated instabilities are localized on the periphery\cite{geyko19}, the match between the fitting function and the data on the periphery is more important than in the interior of the pinch. Third, the experimental data is provided in the interval $r\in[0;0.5]$~cm, which does not define the extension of the fitting function at $r>0$. The main driver of the linear mode\cite{kadomtsev60,angus19} -- the radial logarithmic derivative of the pressure $\frac{r}{P}\frac{dP}{dr}$ -- and, as the result, the stability and spatial location of the mode are fully determined by the choice of the fitting function. The driver can be made arbitrary large by making the density small, therefore, the fitting function should be chosen such that the spatial location of the mode overlaps with the experimental data considerably. This can be achieved by an addition of some small floor to the density function.

Taking into account all the requirements, the following fitting function is introduced
\begin{gather}\label{eq:n_fit}
\frac{n(r)}{n_0} = 0.05 + 0.855\cdot \exp\left(-\frac{r^2}{0.32}-\frac{r^4}{0.0177}\right),
\end{gather}
where $n_0=10^{17}\ \text{cm}^{-3}$, and the radius $r$ is measured in centimeters. Notice that the density has a floor value $0.05$. As it was mentioned, it helps to confine the spatial perturbation of the linear mode at some reasonable radial location. The experimental data with the applied fitting are shown in \Fig{fig:d_fit}.

\begin{figure}
	\centering
	\begin{center}
			\includegraphics[width=0.45\textwidth]{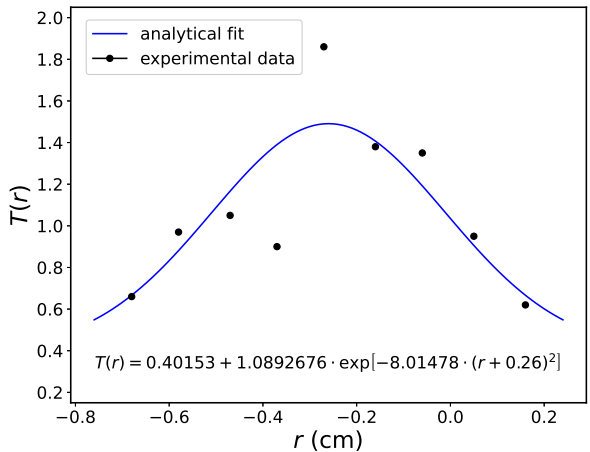}
	\end{center}
	\caption{Least square temperature fitting with a smooth function. Data points are taken from the experimental results\cite{zhang19}. Error bars are ignored and all points are taken with the same weights.}
	\label{fig:T_fit}
\end{figure}

The data for temperature is much more irregular and cannot be fitted well by any smooth function. In this particular realization, the following fitting is used
\begin{gather}\label{eq:T_fit}
\frac{T(r)}{T_0} = 0.4015 + 1.0893\cdot \exp\left[-8.0148(r+0.26)^2\right],
\end{gather}
where $T_0=1.0$~keV, and the radius is again in centimeters. The measured data is centered at $\tilde{r}_0=-0.26$~cm, because the pinch in the experiment moved as a whole from the initial axis when the measurements were performed. This effect is ignored in the present research and the pinch is assumed to be always centered. Thus, the adjusted radial temperature profile reads as
\begin{gather}
\frac{T(r)}{T_0} = 0.4015 + 1.0893\cdot\exp\left[-8.0148 r^2\right],
\end{gather}
and is illustrated in \Fig{fig:T_fit} together with the experimental data.

The self-consistent magnetic field for the equilibrium can be found from the force balance equation $\nabla P = \textbf{J}\times \textbf{B}$, or
\begin{gather}\label{eq:bp}
\frac{\partial P}{\partial r} + \frac{1}{2}\frac{\partial b}{\partial r} + \frac{b}{r} = 0. 
\end{gather} 
Here, pressure $P(r)=P_i+P_e=\left[n_i(r)+n_e(r)\right]T(r)$ is the total plasma pressure, and $b = B^2/4\pi$. The solution of \Eq{eq:bp} is
\begin{gather}\label{eq:bp_1}
b = \frac{b_0}{r^2} - \frac{2}{r^2}\int\limits_0^r \tilde{r}^2\frac{\partial P}{\partial \tilde{r}} d\tilde{r},
\end{gather}
where $b_0/r^2$ is the ``vacuum'' term, i.e. the vacuum magnetic field generated by a linear current at $r=0$, which is absent in the current setup. The second term is of interest, as it describes the magnetic field generated by plasma current. For some analytical profiles (for example, Bennett), the integral in \Eq{eq:bp_1} can be computed exactly. It follows from \Eq{eq:bp} that $\partial P/\partial r$ should be equal to zero at $r=0$, because otherwise the integrated result scales as $r^3$ for small $r$, thus the magnetic field goes as $B\propto \sqrt{r}$, which is an unphysical behavior, as the current density is divergent at $r=0$. This requirement imposes constraints on fitting profiles, as not any arbitrary smooth profile yields to a smooth magnetic field profile. The profiles used for density in \Eq{eq:n_fit} and temperature in \Eq{eq:T_fit} satisfy this constraint.
\begin{figure}
	\centering
	\begin{center}
			\includegraphics[width=0.45\textwidth]{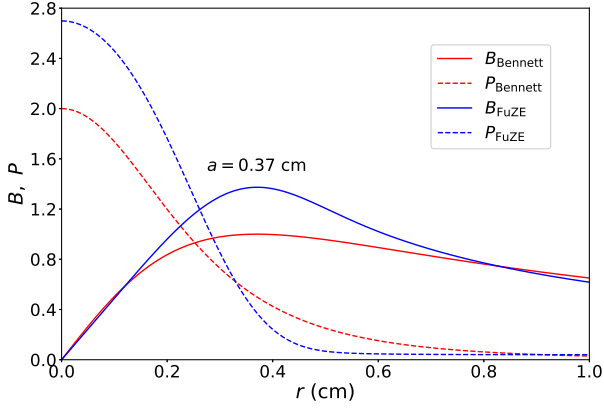}
	\end{center}
	\caption{Comparison of the Bennett and FuZE-like profiles for the same value of the normalization $n_0$ and $T_0$. Both profiles have the maximum value of the magnetic field at $a=0.37$~cm.}
	\label{fig:profiles}
\end{figure}

It is convenient to write \Eq{eq:bp_1} in a dimensionless form where the following variable normalization is used: $B=B_0 \bar{B}$, $n=n_0\bar{n}$, $T=T_0 \bar{T}$, $P=n_0 T_0\bar{P}$, where $n_0$ and $T_0$ were introduced earlier, $B_0^2 = 4\pi n_0 T_0$, and $\xi=r/a$. Notice that in such a normalization, characteristic ion thermal and Alfv\'en velocities are equal $V_{ti}^2=T_0/m_i=V_a^2=B_0^2/(4\pi n_0 m_i)$. This fact is widely used throughout the present paper, as all the velocities are normalized to either thermal or Alfv\'en velocity, which is the same. The solution \Eq{eq:bp_1} reads as
\begin{gather}\label{eq:bp_2}
\bar{B} = \sqrt{\frac{-2}{\xi^2}\int\limits_0^\xi \tilde{\xi}^2\frac{\partial \bar{P}}{\partial \tilde{\xi}} d\tilde{\xi}},
\end{gather}
The integral in \Eq{eq:bp_2} can be computed numerically for any set of points $\xi_i$. Similarly to the Bennett profile, the characteristic radial size of the FuZE-like profile is defined at the position of the maximum magnetic field, in particular in our case it is $a=0.37$~cm. \Fig{fig:profiles} shows the comparison between Bennett and FuZE-like profiles. The latter has much sharper features of the pressure and magnetic field, hence higher logarithmic pressure gradients and greater anticipated linear growth rates.

\subsection{Fluid and mass flow}
\label{sec:flow}

A charged particle in a strong magnetic field moves freely with $v_\parallel$ along the field direction $\textbf{b}$ and orbits around the guiding center with the perpendicular velocity $v_\perp$. The guiding center drifts in the perpendicular direction\cite{bellan06} with $\dot{\textbf{R}}$, so that the guiding center velocity and the parallel acceleration are given by
\begin{gather}\label{eq:GK_v}
\dot{\textbf{R}}_{\alpha} = v_{\parallel} \textbf{b} + c\frac{m_{\alpha} v^2_{\parallel}}{q_{\alpha}B}(\nabla \times \textbf{b})_{\perp} \\ \notag
+ \frac{c}{q_{\alpha}B}\textbf{b}\times(q_{\alpha}\nabla \phi +\mu \nabla B),
\end{gather}
\begin{gather}\label{eq:GK_a}
\dot{v}_{\parallel} =  -\left[\frac{\textbf{b}}{m_{\alpha}}+\frac{c v_{\parallel}}{q_{\alpha}B}(\nabla \times \textbf{b})_{\perp}\right] \cdot (q_{\alpha}\nabla \phi +\mu \nabla B).
\end{gather}
Here, $\alpha$ denotes the species of the particles, ions and electrons in this case, $m_\alpha$ and $q_\alpha$ are the particle mass and charge, $c$ is the speed of light, $\mu=m_\alpha v_\perp^2/(2B)$ is the particle magnetic moment, $\phi$ is the electrostatic potential. In the axisymmetric cylindrical geometry, \Eq{eq:GK_v} simplifies, and the axial component of the guiding center drift velocity reads as 
\begin{gather}\label{eq:vz_d}
\dot{R}_{\alpha,z} =c\frac{m_{\alpha} v^2_{\parallel}}{q_{\alpha}rB} + \frac{c}{q_{\alpha}B}\left(  q_\alpha E_r - \mu \PD{B}{r}\right).
\end{gather}
Assuming a Maxwellian distribution function and making use of \Eq{eq:vz_d}, the axial component of the mean drift velocity can be found
\begin{gather}\label{eq:drift_part}
V_{gc,\alpha} = \frac{1}{n_\alpha}\int\limits_{v^3}\dot{R}_{\alpha,z}f_\alpha(\textbf{x}, \textbf{v})d^3v \\ \notag
=\frac{c}{B}\left[E_r + \frac{T}{q_\alpha r}\left(1-\frac{r}{B}\PD{B}{r}\right)\right],
\end{gather}
where $E_r=-\frac{\partial \phi}{\partial r}$ is the radial component of electric field. The first term in the right hand side of \Eq{eq:drift_part} is the $E{\times}B$ drift, and the last term is the combination of magnetic drifts. 
For a given pressure and plasmas flow profiles, the electric field is uniquely defined\cite{bellan06} from the following equation

\begin{figure}
	\centering
	\begin{center}
			\includegraphics[width=0.45\textwidth]{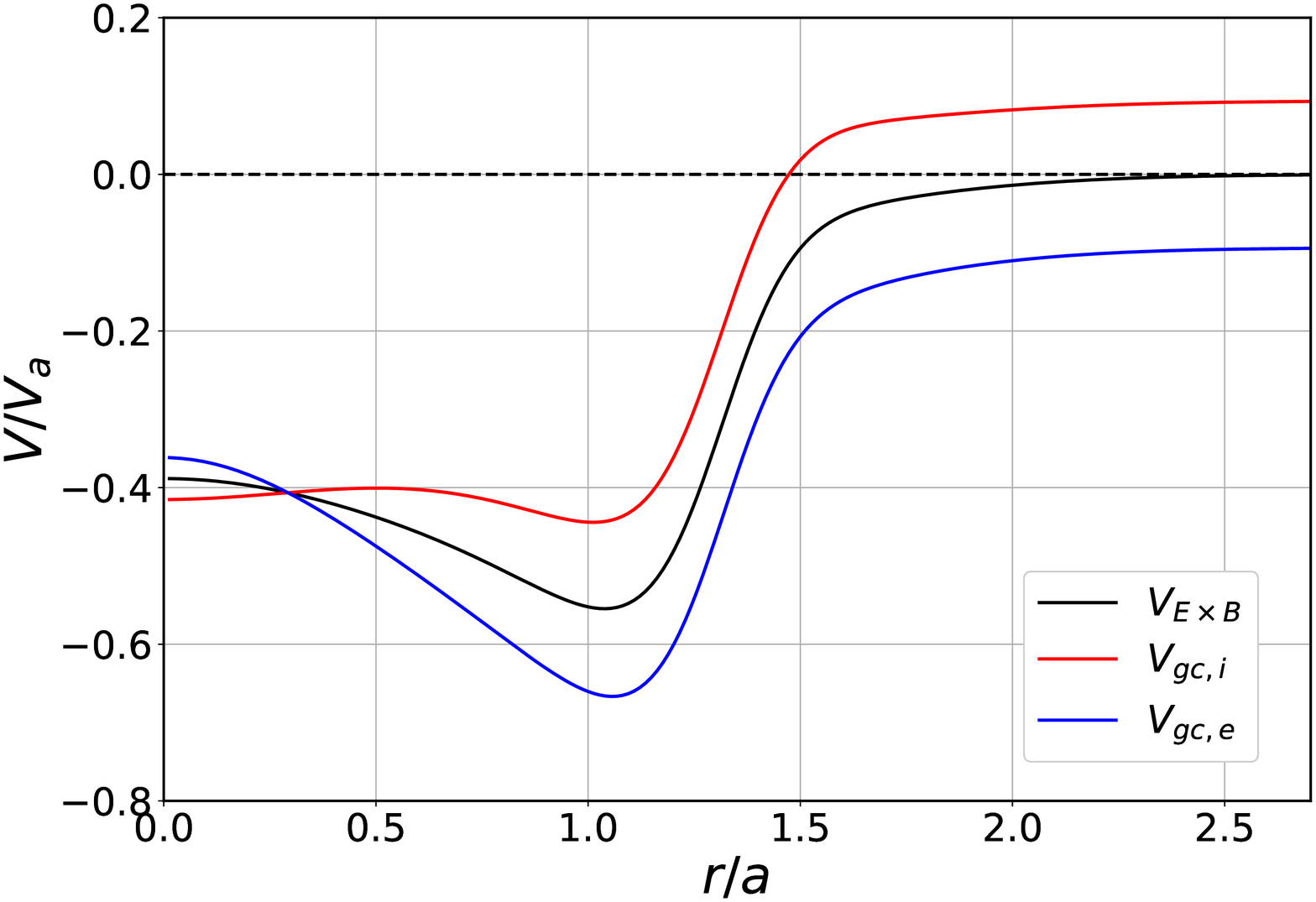}
	\end{center}
	\caption{Axial components of the drift velocities for FuZE-like pinch equilibrium: $E{\times}B$ drift compared to the total ion and electron drifts. All velocities are normalized to the ion thermal velocity $V_{ti}$.}
	\label{fig:drifts}
\end{figure}

\begin{gather}\label{eq:fluid_gk_full}
\frac{\textbf{v}_{i,e}}{c} = \frac{\textbf{E}\times \textbf{B}}{B^2} \mp \frac{\nabla P_{i,e}\times \textbf{B}}{qn_{i,e}B^2},
\end{gather}
where $q$ is the elementary charge, and singly ionized ions $q_i=-q_e=q$ are considered for simplicity. \Eq{eq:fluid_gk_full} is the expression for the fluid velocities of ions and electrons $\textbf{v}_{i,e}$ in terms of the $E{\times}B$ and diamagnetic drifts. For stationary ions $\textbf{v}_i=0$, the difference between the $E{\times}B$ and total drift velocity of ions and electrons \Eq{eq:drift_part} is illustrated in \Fig{fig:drifts}, where the parameters are those from the FuZE-like pinch.

\begin{figure}
	\centering
	\begin{center}
			\includegraphics[width=0.45\textwidth]{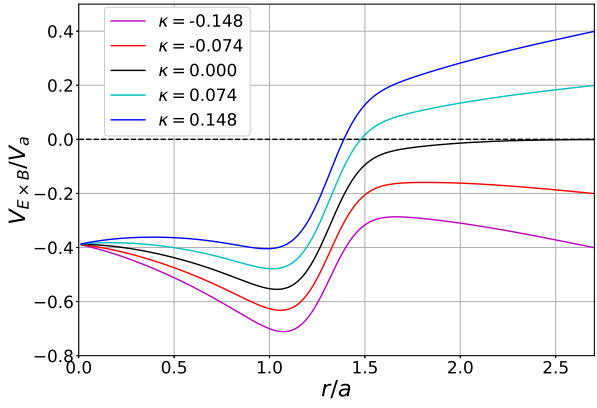}
	\end{center}
	\caption{Normalized $E{\times}B$ velocity as a function of radius for different values of shear parameter $\kappa$.}
	\label{fig:ExB}
\end{figure}

\begin{figure}
	\centering
	\begin{center}
			\includegraphics[width=0.45\textwidth]{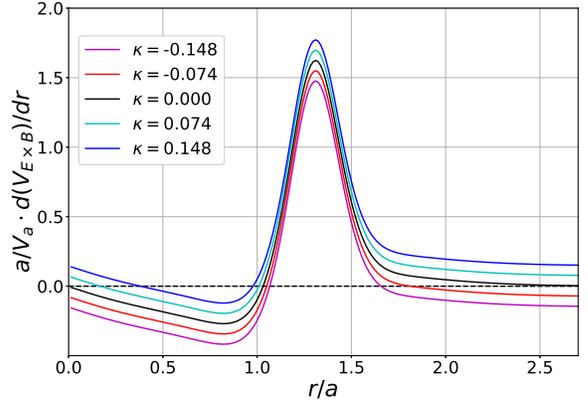}
	\end{center}
	\caption{Radial derivative of the normalized $E{\times}B$ velocity as a function of radius for different values of shear parameter $\kappa$. }
	\label{fig:ExB_sh}
\end{figure}

For an axisymmetric cylindrical geometry, \Eq{eq:fluid_gk_full} simplifies to
\begin{gather}\label{eq:fluid_gk}
\frac{\hat{z}\cdot\textbf{v}_{i,e}}{c} = \frac{E_r}{B} \mp \frac{\partial P_{i,e}}{\partial r}\frac{1}{qn_{i,e}B}.
\end{gather}
For subsonic flows, ion fluid velocity $v_{zi}$ is a fraction of the thermal velocity $V_{ti}=\sqrt{T_i/m_i}$, thus the first term in \Eq{eq:fluid_gk} scales as $V_{ti}/c$. The second term on the right hand side of \Eq{eq:fluid_gk} scales as
\begin{gather}\label{eq:fluid_gk1}
\frac{\partial P_{i,e}}{\partial r}\frac{1}{qn_{i,e}B} \propto \frac{nT}{aqnB}\propto \frac{V_{ti}}{c}\frac{V_{ti}}{\omega_{ci}a}\propto\epsilon\frac{V_{ti}}{c},
\end{gather} 
where $a$ is the characteristic radial scale of the pinch, $\omega_{ci}=qB/(m_i c)$ is the ion cyclotron frequency and the magnetization parameter $\epsilon=\rho_i/a=V_{ti}/(a \omega_{ci})$. In strongly magnetized plasmas, $\epsilon\ll 1$, thus the fluid velocity is approximately equal to the drift velocity. Therefore, a fluid shear flow is required for the existence of the $E{\times}B$ velocity shear and vice versa. This assumption is violated in the case of FuZE-like profiles. First, the parameter $\epsilon$ is not vanishingly small, but instead is equal to $0.1{-}0.2$ depending at what location inside the pinch it is measured. Second, the profile itself has very sharp gradients, so the last term in \Eq{eq:fluid_gk1} becomes comparable to the other terms, therefore strong $E{\times}B$ velocity shear can be present even in the absence of the fluid flow. In this work, we consider linear shear flow, $V_{sh}=\kappa V_a r/a$, and the parameter $\kappa$ is called a \textit{shear parameter}. \Fig{fig:ExB} demonstrates how the $E{\times}B$ velocity depends on the radius for the FuZE-like profile for 5 different values $\kappa$. \Fig{fig:ExB_sh} shows the radial derivative of the $E{\times}B$ velocity for the same values of $\kappa$. There is a very noticeable spike of the derivative at $r/a\approx 1.3$, where embedded shear value corresponds to $\kappa\approx 1.5$.

The conjecture is made here, that in collisionless plasmas, the instability dynamics is predominantly determined by the guiding center velocity, as the motion of every individual particle is only determined by drifts. As it follows from \Fig{fig:drifts}, the difference between the guiding center velocity and the $E{\times}B$ velocity is small, especially, when the radial derivative is considered. Assuming the conjecture is true, the two conclusions follow. First, if there is an embedded shear of the drift motion in the system, the amount of the fluid flow shear required to change the system behavior should be at least comparable to the intrinsic value of the guiding center shear. Second, if any stabilization via sheared flow exists, then it should apply to the embedded guiding center velocity shear as well, therefore addition of an extra fluid flow shear can be both stabilizing or destabilizing, depending on how the fluid shear is related to the embedded one. For example, for the FuZE-like profile, positive $\kappa$ increases the radial derivative of the $E{\times}B$ velocity, while negative $\kappa$ decreases (See \Fig{fig:ExB_sh}), hence, more suppression of the linear mode should be anticipated for positive $\kappa$.

Returning to the comparison of different pinch profiles, notice a peculiarity of the commonly used Bennett profile. The diamagnetic term in \Eq{eq:fluid_gk1} is independent of $r$ for any values of the parameter $\epsilon$, which means that the Bennett profile is unique, as no embedded shear of the $E{\times}B$ drift is present. Moreover, even the total guiding center drift, including the magnetic field corrections in \Eq{eq:drift_part}, does not depend on $r$ either. As a consequence of that, simulations based on the Bennett profile do not provide a general picture, as they are lacking important physics related to the embedded guiding center drift shear.

\section{Gyrokinetic simulations}
\label{sec:gk_sec}

The gyrokinetic simulations are performed with the high-order finite volume code COGENT\cite{dorf13}. The code numerically solves the following equation for the gyro-distribution function $f_\alpha$
\begin{gather}
\frac{\partial}{\partial t}\left(B_{\parallel\alpha}^* f_\alpha\right)+\nabla_\textbf{R}\cdot\left(\dot{\textbf{R}}_\alpha B_{\parallel\alpha}^* f_\alpha\right)+\frac{\partial}{\partial v_\parallel}\left(\dot{v}_{\parallel\alpha}B_{\parallel\alpha}^* f_\alpha\right)=0,
\end{gather}
where $\nabla_\textbf{R}$ denotes a differential operator with respect to the guiding center coordinates, $\textbf{b}=\textbf{B}/B$ is the unit vector in the direction of the magnetic field, $B_{\parallel\alpha}=\textbf{B}_{\parallel\alpha}\cdot\textbf{\textbf{b}}$, and
\begin{gather}
\textbf{B}_{\parallel\alpha} = \textbf{B} + \frac{m_\alpha c v_\parallel}{q_\alpha}\nabla\times \textbf{b}.
\end{gather}
The guiding center drift and parallel acceleration are given by \Eq{eq:GK_v} and \Eq{eq:GK_a} respectively. The simulations are performed in the 2D cylindrical configuration space $(r,z)$ with angular symmetry assumed, and the 2D velocity space $(v_\parallel,\mu)$. The simulation domain has radial boundaries at $r=0.18$ and $r=0.82$~cm, with 64 cells in the radial direction and 32 cells in the axial direction. The domain and the density fitting function floor in \Eq{eq:n_fit} are chosen such that the spatial mode is localized away from the external radial boundary at $r=R_\text{max}$ in order to eliminate any possible boundary stabilization effects. In the axial direction, only one full wavelength of the mode is seeded. The domain spans from $z=0$ to $z=\lambda$, where different values of $\lambda$ are tested (from $0.16$ to $2.56$~cm). Periodic boundary condition are applied in the axial direction. Dirichlet boundary condition at $r=R_\text{max}$ and Neumann at $r=R_\text{min}$ are used for the potential. The linear growth rate is measured as a function of the wave vector $k_z = 2\pi/\lambda$ for different values of $\kappa$. More details of the simulations and the methodology of the measurements are provided in the previous work of Geyko \textit{et al}\cite{geyko19}.

\begin{figure}
	\centering
	\begin{center}
			\includegraphics[width=0.45\textwidth]{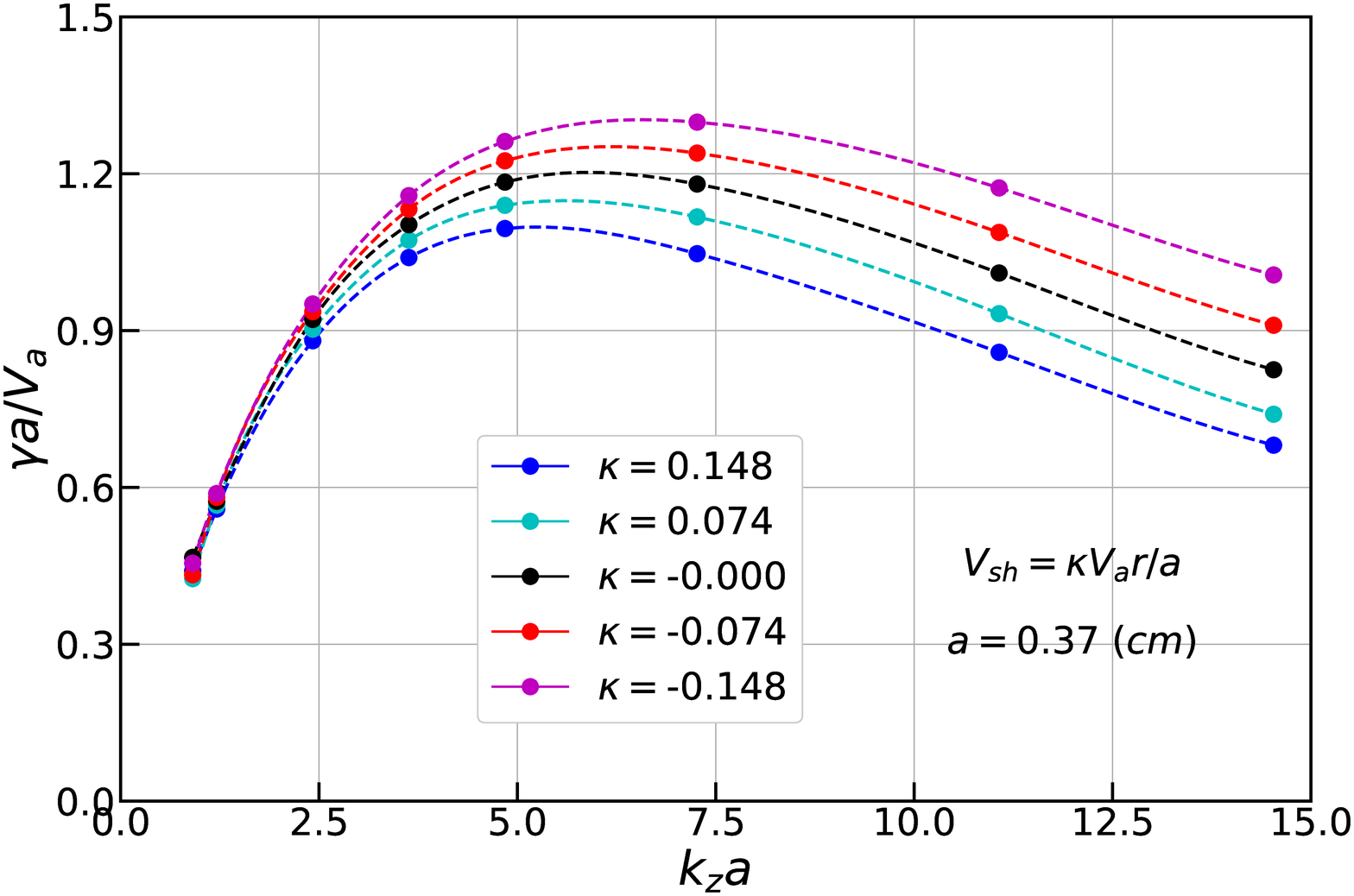}
	\end{center}
	\caption{Normalized growth rate of the linear $m=0$ mode for the FuZE-like profile as a function of the axial wave vector $k_z a$. 5 different values of the shear parameter are considered.}
	\label{fig:gr_fuze}
\end{figure}

Linear growth rates obtained from the simulations are shown in \Fig{fig:gr_fuze} and the main observations are the following. First, the growth rate curve $\gamma(k_z)$ has the same shape as the one obtained for the Bennett case\cite{geyko19}, namely it has a rollover at high $k_z$ part of the spectra. Second, the main difference now is that the problem has become shear direction dependent, and a moderate fluid shear ($\kappa\sim 0.2$) can play even a destabilizing role, if the direction is not properly chosen. Finally, the growth rate dependence on $\kappa$ is quite weak, and no stabilization is observed for $\kappa$ on the order of a fraction of unity.

All these observations are in agreement with the conjecture in \Sec{sec:flow}. Indeed, if the guiding center shear is what determines linear mode stability, then slightly changed by the fluid flow shear (according to \Fig{fig:ExB_sh}) it does not have a significant affect on the growth rate. If stronger shear is required for better stability, then according to \Fig{fig:gr_fuze} positive fluid flow shear $\kappa>0$ should lead to more mode stabilization and negative shear $\kappa<0$ should do the opposite, exactly what is demonstrated in \Fig{fig:gr_fuze}. It also follows from the conjecture that for the fluid flow shear to be comparable to the intrinsic $E{\times}B$ velocity shear, it should be at least $\kappa\approx 1.5$, i.e. supersonic shear, which is not observed in the experiments.

\begin{figure}
	\centering
	\begin{center}
			\includegraphics[width=0.45\textwidth]{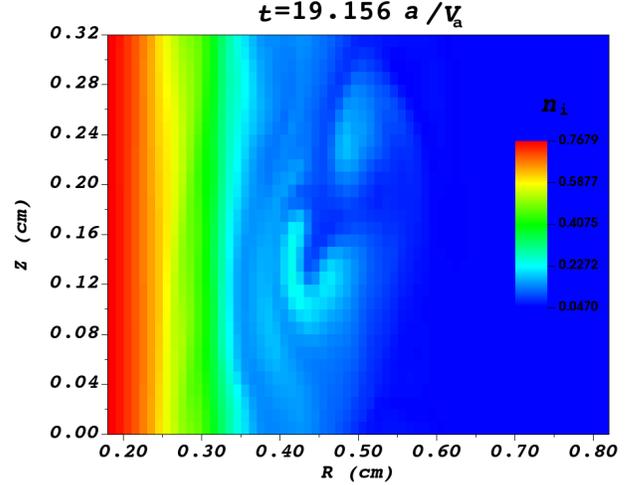}
	\end{center}
	\caption{Density plot for the nonlinear evolution of the $m=0$, $k_z a = 7.26$ mode for the the FuZE-like profile. The perturbations are localized on the periphery.}
	\label{fig:fuze_n32}
\end{figure}

\begin{figure}
	\centering
	\begin{center}
			\includegraphics[width=0.45\textwidth]{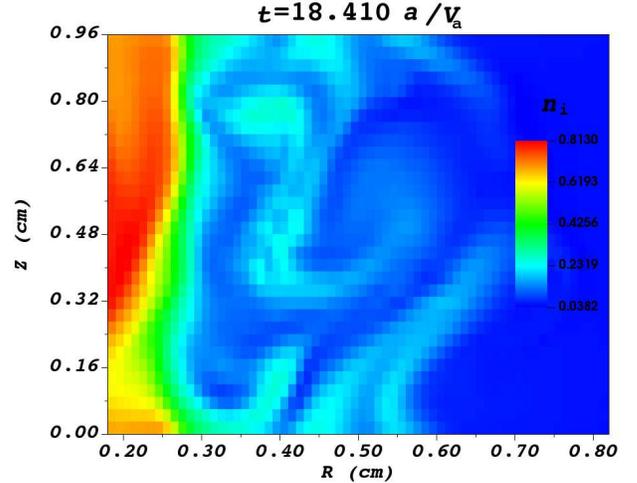}
	\end{center}
	\caption{Density plot for the nonlinear evolution of the $m=0$, $k_z a = 2.42$ mode for the the FuZE-like profile. The perturbations are spread into the interior of the pinch.}
	\label{fig:fuze_n96}
\end{figure}

As no linear mode mitigation is revealed in the numerical simulations, the source of Z-pinch stabilization in the experiments remains unclear. To address this question, we look at the nonlinear evolution of the system simulated for different values of the domain size in the axial direction, namely, $\lambda = 0.32$ and $\lambda=0.96$~cm. In both cases, the linear modes with the corresponding wavelengths are initially seeded and evolve to the nonlinear stage. Strictly speaking, a nonlinear evolution assumes the presence of all the modes supported by the system, therefore artificial limitation on the domain size cuts off the long wavelength part of the spectra and does not provide a complete physical picture. The case of the small domain ($\lambda = 0.32$) is nevertheless studied both for academic purposes and also for comparison with the similar evolution of the turbulence in the Bennett profile case. \Fig{fig:fuze_n32} and \Fig{fig:fuze_n96} show the density plots of the nonlinear evolution for $\lambda = 0.32$ and $\lambda=0.96$~cm modes respectively. The main difference is that the perturbations of the short wavelength mode are located on the periphery and do not propagate into the interior of the pinch. The bulk of the pinch is then not perturbed, and therefore a nonlinear stability can be claimed. It is not the case for the long wavelength modes though, as illustrated in \Fig{fig:fuze_n96}. The perturbations reach the inner boundary of the simulation domain, and in principle, they can possibly spread into the interior of the pinch.

\begin{figure}
	\centering
	\begin{center}
			\includegraphics[width=0.45\textwidth]{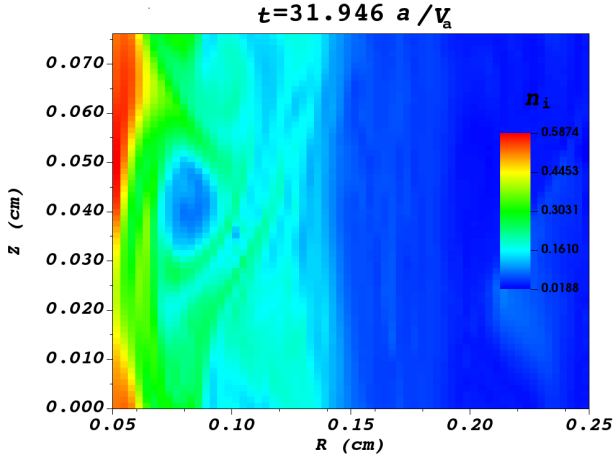}
	\end{center}
	\caption{Density plot for the nonlinear evolution of the $m=0$, $k_z a = 7.5$ mode for the the Bennett profile. The perturbations are spread into the interior of the pinch.}
	\label{fig:bennett}
\end{figure}

It is interesting to compare the nonlinear evolution of the same wavelength for Bennett and FuZE-like profiles, in particular, a case of nonlinear stabilization of the FuZE-like pinch. Two similar normalized wavelengths are chosen: $k_z a=7.26$ for FuZE-like and $k_z a=7.5$ for Bennett cases. \Fig{fig:bennett} demonstrates that nonlinear perturbations of the mode for the Bennett pinch are not confined on the periphery and instead propagate to the inner boundary of the domain. Since no fluid flow shear is involved ($\kappa=0$) in both simulations, this phenomenon is purely due to the pinch profile shape.

\section{Extended MHD simulations}
\label{sec:mhd_sec}

The present gyrokinetic model suffers from the absence of electromagnetic effects that are especially important for long wavelength modes\cite{geyko19}, and from the limitations on the drift velocity, which should not exceed sonic speed. These issues can be addressed in the extended-MHD model, also implemented in the COGENT code. The MHD model simulates the following equations.
The continuity equation reads as
\begin{gather}\label{eq:rho_f}
\PD{\rho}{t}+\nabla\cdot \textbf{m} = 0,
\end{gather}
where $\rho\approx m_i n_i$ is the fluid density and $\textbf{m}= \rho \textbf{u}$ is the momentum density with $\textbf{u}\approx \textbf{v}_i$ as the electron-ion mass ratio is very small. The momentum density is governed by the MHD equation of motion
\begin{gather}
\PD{\textbf{m}}{t} + \nabla \cdot \left( \textbf{m}\textbf{u} + \textbf{I}P+\pmb{\pi} \right) = \frac{\textbf{J}\times \textbf{B}}{c}.
\end{gather}
Here, $\pmb{\pi}$ is the gyro-viscous pressure tensor, based on Braginskii's formulation. The ion and electron energy densities are defined as
\begin{gather}
\varepsilon_i=\frac{1}{2}\rho u^2+\frac{P_i}{\gamma-1}, \\ \notag
\varepsilon_e=\frac{P_e}{\gamma-1}.
\end{gather}
The equations for $\varepsilon_i$ and $\varepsilon_e$ include diamagnetic heat fluxes $\textbf{q}_i$ and $\textbf{q}_e$, 
\begin{gather}\label{eq:e_i}
\PD{\varepsilon_{i}}{t}+\nabla\cdot\left[\textbf{u}(\varepsilon_i+P_i)+\textbf{u}\pmb{\pi}+\textbf{q}_i\right]=
qn\textbf{u}\cdot\textbf{E}+Q_{ie},
\end{gather}
\begin{gather}\label{eq:e_e}
\PD{\varepsilon_e}{t}+\nabla\cdot\left[\textbf{u}_e(\varepsilon_e+P_e)+\textbf{q}_e\right]=
-qn\textbf{u}_e\cdot\textbf{E}-Q_{ie}.
\end{gather}
In Eqs.~(\ref{eq:e_i},\ref{eq:e_e}), $Q_{ie}$ is the ion-electron heat exchange, and $\textbf{u}_e$ is the electron velocity, which is given by
\begin{gather}\label{eq:u_e}
\textbf{u}_e=\textbf{u}-\frac{\textbf{J}}{qn}.
\end{gather}
Equations~(\ref{eq:rho_f}-\ref{eq:u_e}) are coupled with the generalized Ohm's law
\begin{gather}
m_e\PD{\textbf{J}}{t} =q^2n\left[\textbf{E}+\frac{\textbf{u}\times \textbf{B}}{c} \right. \\ \notag
\left. -\frac{1}{qn}\left(\frac{\textbf{J}\times\textbf{B}}{c}-\nabla P_e\right)\right],
\end{gather}
Ampere's law 
\begin{gather}
\PD{\textbf{E}}{t}=c\nabla\times\textbf{B}-4\pi\textbf{J},
\end{gather}
and Faraday's laws
\begin{gather}
\PD{\textbf{B}}{t}+c\nabla\times\textbf{E}=0.
\end{gather}
The diamagnetic heat flux terms $\textbf{q}_{i,e}$ are responsible for the existence of certain drift modes in the MHD model, such as the entropy mode studied by different authors\cite{kadomtsev60, simakov01, ricci06prl, angus19}. The model equations are identical to those used in the work of Angus \textit{et al}\cite{angus19} with the exception that gyro-viscosity is included here in this work. More details of the model and MHD simulations can be found in Ref.~\cite{angus19}.

\begin{figure}
	\centering
	\begin{center}
			\includegraphics[width=0.45\textwidth]{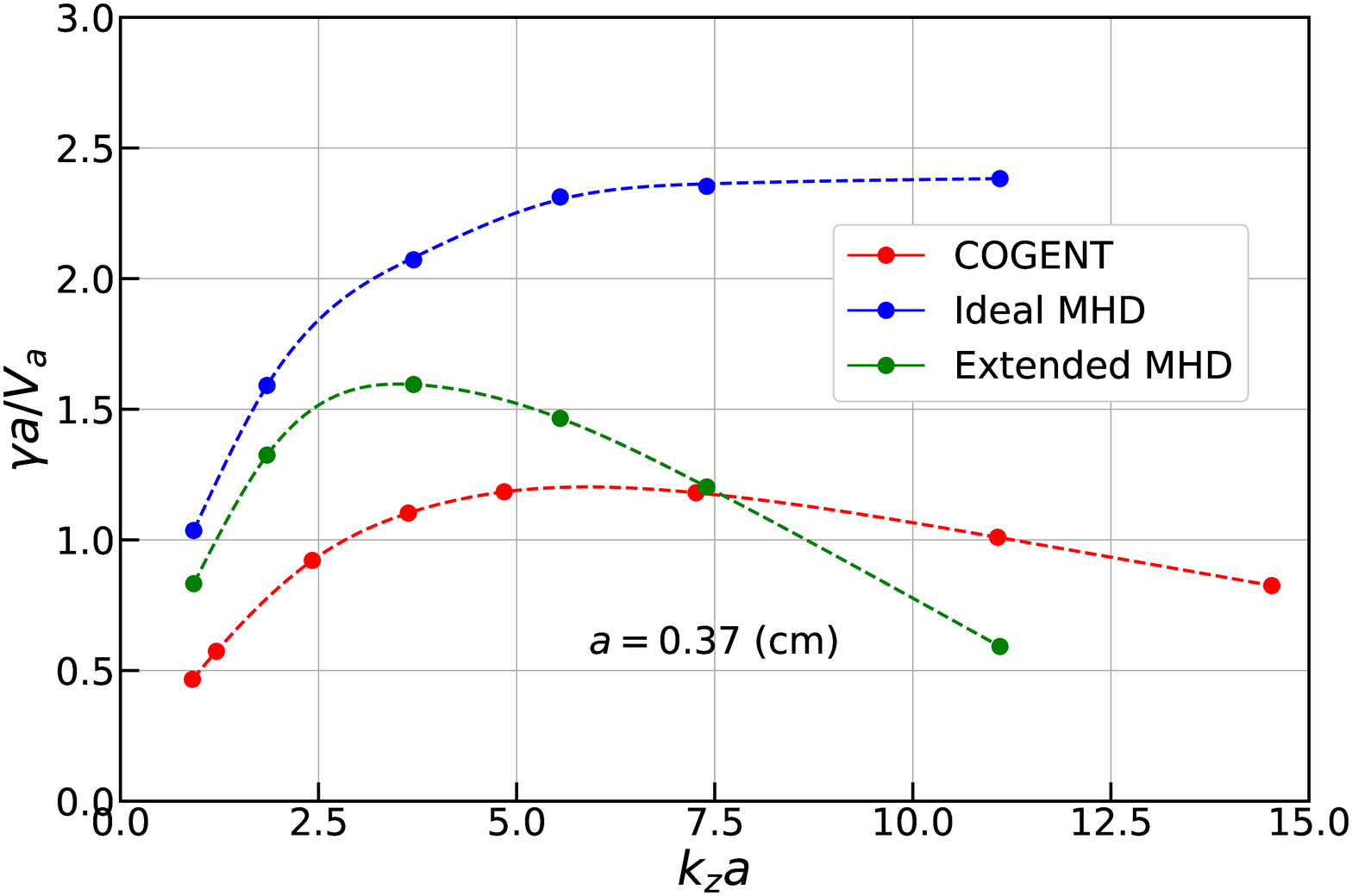}
	\end{center}
	\caption{Comparison of shear-less growth rate curves. Red: gyrokinetic COGENT, blue: ideal MHD, green: extended MHD.}
	\label{fig:gk_mhd}
\end{figure}

The extended-MHD model is tested and compared to the electrostatic gyrokinetic one for the FuZE-like pinch profile. The growth rate of a linear $m=0$ mode is found for different values of the normalized axial wave vector $k_z a$, and the results are shown in \Fig{fig:gk_mhd}. The growth rate roll-over effect at the high-$k$ part of the spectra is reproduced and it is in a reasonable match with the one from the gyrokinetic simulations\cite{geyko19}. This effect appears in the MHD model due to the gyro-viscous pressure tensor $\pmb{\pi}$. Interestingly, the growth rate obtained by the ideal-MHD simulations is considerably greater than one from the extended-MHD, which suggests that FRL effects play an important role in the linear mode stabilization. Furthermore, this circumstance demonstrates the significance of the pinch profile, as the difference between the ideal-MHD, gyrokinetic and fully kinetic simulations are much less for the Bennett profile (Fig.~(3) from the work of Geyko \textit{et al}\cite{geyko19}).

The nonlinear mechanism of the stabilization is also verified via the extended-MHD simulations. The wavelengths of the modes are chosen differently, yet of the same order that used in the gyrokinetic simulations, namely, $k_z a = 11.1$ for the short wavelength and $k_z a = 3.7$ for the long wavelength mode. The results of the MHD simulations are consistent with the gyrokinetic ones: short wavelength perturbations nonlinearly saturate on the periphery, while long wavelengths perturbations penetrate into the interior and completely disrupt the pinch.

\begin{figure}
	\centering
	\begin{center}
			\includegraphics[width=0.45\textwidth]{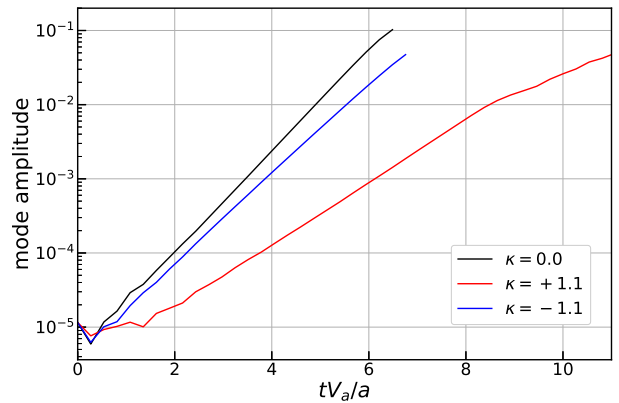}
	\end{center}
	\caption{Growth of the linear mode $k_z a =3.7$ for three values of $\kappa$: -1.1, 0.0 and 1.1. In all three cases, the mode is unstable.}
	\label{fig:mhd_kappa}
\end{figure}

Finally, large fluid flow shears $\kappa=\pm 1.1$ are tested, where the variation of the flow over the pinch radius is greater than the Alfv\'en speed. Even this unrealistic flow shear is shown to be insufficient to fully mitigate linear modes. \Fig{fig:mhd_kappa} is a logarithmic plot of the perturbation amplitude of $k_z a =3.7$ mode as a function of time. The mode is unstable for all three values of $\kappa$, and the only difference is the growth rate, which is nearly unchanged for $\kappa=-1.1$ and noticeably lower for $\kappa=1.1$. This observation is consistent with the previously mentioned conclusion that, in the case of FuZE-like profile, the amount of shear required for suppression of linear instabilities should be at least comparable to the amount of embedded guiding center drift shear. For the parameters of the problem, it is approximately $\kappa=1.5$ (See \Fig{fig:ExB_sh}). This requirement is only necessary but not sufficient, which was demonstrated in the simulations.

\section{Profile flattening stabilization}
\label{sec:flatten}

\begin{figure}
	\centering
	\begin{center}
			\includegraphics[width=0.45\textwidth]{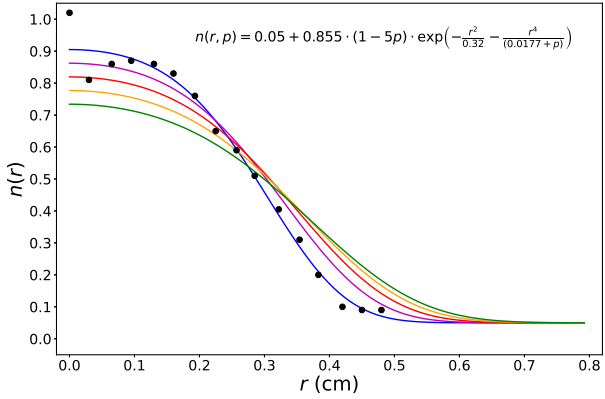}
	\end{center}
	\caption{Possible relaxation of the density profile fitting curve for 5 different values of the parameter $p$ in \Eq{eq:flat_n}.}
	\label{fig:flatten}
\end{figure}

As it was shown in \Sec{sec:gk_sec}, nonlinear pinch stabilization can be achieved for short wavelengths, and the stabilization mechanism is the pinch profile itself rather than fluid shear flow. In this section, the nonlinear stabilization is investigated in more details by arbitrarily relaxing the density profile while maintaining it `close' to the experimental data. For example, one can argue that experimental data varies depending on the axial location of the measurements and is obtained with some errors\cite{zhang19}. Thus, a family of fitting profiles is considered, for example,
\begin{gather}\label{eq:flat_n}
n(r,p) = 0.05 + 0.855\cdot p \\ \notag
+ 0.855\cdot (1-p)\cdot \exp\left(-\frac{r^2}{0.32}-\frac{r^4}{(0.0177+p)}\right)
\end{gather}
where $p$ is a \textit{flattening parameter}. The dependence of the density curve on $p$ is shown in \Fig{fig:flatten}. The temperature is kept constant and equal to $T=1.0$~keV for simplicity. As $p$ increases, the match between the experimental data and the fitting curve becomes worse, so there is no point to consider $p>0.04$ if the experimental profile is implied. While the model is indeed inaccurate and should not be considered as a rigorous analysis, it is illustrative and suitable for better understanding of the stabilization phenomena. 

Simulations for different values of the parameter $p$ and different wavelengths have been performed. The first observation is that linear mode stabilization cannot be achieved via profile flattening. The growth rate decreases as $p$ gets larger, yet even for $p=0.04$, which is very far from the experimental data and the normalized growth rate of the most unstable mode ($\lambda=0.32$~cm) only decreases from 1.05 to 0.6. 
 
\begin{figure}
	\centering
	\begin{center}
			\includegraphics[width=0.45\textwidth]{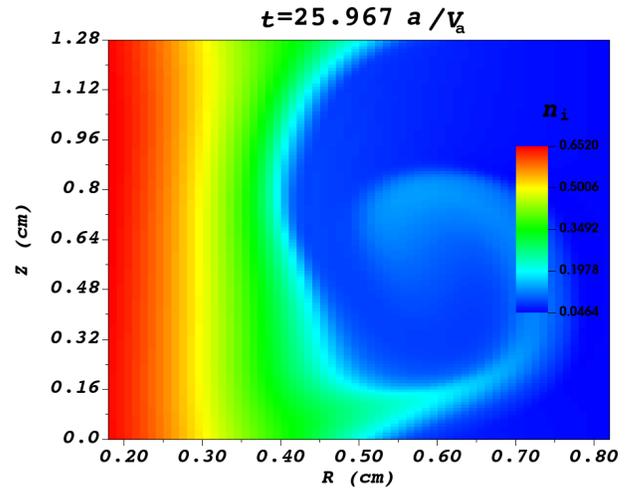}
	\end{center}
	\caption{Density plot for the nonlinear evolution of $\lambda=0.96$~cm mode for the case of $p=0.04$ test profile. The perturbations are saturated on the periphery.}
	\label{fig:n_96_flat}
\end{figure}

The nonlinear behavior is, however, significantly different. The nonlinear perturbation shifts from the interior of the pinch to the periphery, very similar as it was observed for short wavelength modes for the FuZE-like pinch in \Fig{fig:fuze_n32}. A density plot of the nonlinear evolution of $\lambda=1.28$~cm mode is demonstrated in \Fig{fig:n_96_flat}. This mode is shown completely unstable in both gyrokinetic and extended-MHD simulation. For the test profile with p=0.04, all of the perturbations are located at $r>0.4$, which is at the outer periphery of the pinch. Thus, the interior remains unperturbed and global pinch stabilization can be claimed.

The flattening of the profile basically pushes the instabilities to the periphery and exploits the nonlinear stabilization mechanism described for the short wavelength modes. Nevertheless, it remains arguable how robust such the mechanism can be as the required modifications to the profiles can be so large that they might not adequately represent the experimental data.

\section{Disscussion and Conclusion}
\label{sec:concl}

The stability properties of $m=0$ modes in a FuZE-like type of a Z-pinch have been studied in the present work. The pinch profile is shown to play an important role in both the growth rate of linear modes and stabilization possibilities via axially sheared fluid flow. The key difference of any realistic profile, including the FuZE-like one, from the model Bennett profile is that the particle guiding center velocity has an intrinsic embedded shear, even for a zero fluid flow. Unlike the Bennett case, the presence of a fluid flow shear does not necessarily lead to the mode mitigation and the reduction of the growth rate, but can also make the system even more unstable. The intrinsic guiding center flow, however, for the parameters of the FuZE experiment, has quite a pronounced (up to Alfv\'en speed over the pinch radius) local shear, thus, no moderate (sub-Alfv\'enic) flow shear is sufficient to change the instability behavior drastically.

These conjectures have been confirmed by numerical simulations performed with the COGENT code. Both electrostatic gyrokinetic and extended-MHD models agreed that the growth rate of the linear $m=0$ mode is not considerably affected by a subsonic fluid shear flow. The changes of the growth rate corresponding to the sign of the applied shear found to be consistent with the guiding center drift picture. The nonlinear stabilization of the short wavelength modes ($ka\ge 7.5$ in gyrokinetic $ka\ge 11.1$ in extended-MHD simulations) has been observed. The mechanism of such stabilization is due to nonlinear saturation of the modes on the pinch periphery such that the interior of the pinch remains unperturbed. The global pinch stability has not been achieved, as the mechanism is unable to stabilize long wavelength modes.

The results presented in this work do not support the conjecture that Z-pinch stabilization observed in some experiments is due to a sheared axial flow of the plasma. It is shown that if a realistic pinch profile is considered, no sub-Alfv\'enic fluid flow shear is sufficient for stabilization of linear modes. Relaxation of  pressure gradients, described in \Sec{sec:flatten}, is not sufficient either. A possible explanation is that the linear modes are always unstable and the global pinch stability is achieved by a combination of nonlinear saturation of the modes and finite Larmor radius effects. In order to proceed and investigate this phenomenon thoroughly, more detailed computational models are needed. To that end, higher order FLR effects are developed, included and being tested in the COGENT code, as well as further improvements of the extended-MHD model are being done.

\section{Acknowledgments}
This work was performed under the auspices of US DOE by LLNL under Contract DE-AC52-07NA27344 and was supported by LLNL-LDRD under Project No. 18-ERD-007.

\bibliographystyle{aipnum4-1}
\bibliography{bibliography_m}

\begin{thebibliography}{24}%
\makeatletter
\providecommand \@ifxundefined [1]{%
 \@ifx{#1\undefined}
}%
\providecommand \@ifnum [1]{%
 \ifnum #1\expandafter \@firstoftwo
 \else \expandafter \@secondoftwo
 \fi
}%
\providecommand \@ifx [1]{%
 \ifx #1\expandafter \@firstoftwo
 \else \expandafter \@secondoftwo
 \fi
}%
\providecommand \natexlab [1]{#1}%
\providecommand \enquote  [1]{``#1''}%
\providecommand \bibnamefont  [1]{#1}%
\providecommand \bibfnamefont [1]{#1}%
\providecommand \citenamefont [1]{#1}%
\providecommand \href@noop [0]{\@secondoftwo}%
\providecommand \href [0]{\begingroup \@sanitize@url \@href}%
\providecommand \@href[1]{\@@startlink{#1}\@@href}%
\providecommand \@@href[1]{\endgroup#1\@@endlink}%
\providecommand \@sanitize@url [0]{\catcode `\\12\catcode `\$12\catcode
  `\&12\catcode `\#12\catcode `\^12\catcode `\_12\catcode `\%12\relax}%
\providecommand \@@startlink[1]{}%
\providecommand \@@endlink[0]{}%
\providecommand \url  [0]{\begingroup\@sanitize@url \@url }%
\providecommand \@url [1]{\endgroup\@href {#1}{\urlprefix }}%
\providecommand \urlprefix  [0]{URL }%
\providecommand \Eprint [0]{\href }%
\providecommand \doibase [0]{http://dx.doi.org/}%
\providecommand \selectlanguage [0]{\@gobble}%
\providecommand \bibinfo  [0]{\@secondoftwo}%
\providecommand \bibfield  [0]{\@secondoftwo}%
\providecommand \translation [1]{[#1]}%
\providecommand \BibitemOpen [0]{}%
\providecommand \bibitemStop [0]{}%
\providecommand \bibitemNoStop [0]{.\EOS\space}%
\providecommand \EOS [0]{\spacefactor3000\relax}%
\providecommand \BibitemShut  [1]{\csname bibitem#1\endcsname}%
\let\auto@bib@innerbib\@empty
\bibitem [{\citenamefont {Reynolds}\ and\ \citenamefont
  {Craggs}(1952)}]{reynolds52}%
  \BibitemOpen
  \bibfield  {author} {\bibinfo {author} {\bibfnamefont {P.}~\bibnamefont
  {Reynolds}}\ and\ \bibinfo {author} {\bibfnamefont {J.~D.}\ \bibnamefont
  {Craggs}},\ }\href@noop {} {\bibfield  {journal} {\bibinfo  {journal}
  {Philos. Mag.}\ }\textbf {\bibinfo {volume} {43}} (\bibinfo {year}
  {1952})}\BibitemShut {NoStop}%
\bibitem [{\citenamefont {Kurchatov}(1957)}]{kurchatov57}%
  \BibitemOpen
  \bibfield  {author} {\bibinfo {author} {\bibfnamefont {I.~V.}\ \bibnamefont
  {Kurchatov}},\ }\href@noop {} {\bibfield  {journal} {\bibinfo  {journal} {J.
  Nucl. Eng.}\ }\textbf {\bibinfo {volume} {4}} (\bibinfo {year}
  {1957})}\BibitemShut {NoStop}%
\bibitem [{\citenamefont {Kadomtsev}(1960)}]{kadomtsev60}%
  \BibitemOpen
  \bibfield  {author} {\bibinfo {author} {\bibfnamefont {B.~B.}\ \bibnamefont
  {Kadomtsev}},\ }\href@noop {} {\bibfield  {journal} {\bibinfo  {journal}
  {JETP}\ }\textbf {\bibinfo {volume} {10}},\ \bibinfo {pages} {780} (\bibinfo
  {year} {1960})}\BibitemShut {NoStop}%
\bibitem [{\citenamefont {Ricci}\ \emph {et~al.}(2006)\citenamefont {Ricci},
  \citenamefont {Rogers}, \citenamefont {Dorland},\ and\ \citenamefont
  {Barnes}}]{ricci06pop}%
  \BibitemOpen
  \bibfield  {author} {\bibinfo {author} {\bibfnamefont {P.}~\bibnamefont
  {Ricci}}, \bibinfo {author} {\bibfnamefont {B.~N.}\ \bibnamefont {Rogers}},
  \bibinfo {author} {\bibfnamefont {W.}~\bibnamefont {Dorland}}, \ and\
  \bibinfo {author} {\bibfnamefont {M.}~\bibnamefont {Barnes}},\ }\href@noop {}
  {\bibfield  {journal} {\bibinfo  {journal} {Phys. Plasmas}\ }\textbf
  {\bibinfo {volume} {13}} (\bibinfo {year} {2006})}\BibitemShut {NoStop}%
\bibitem [{\citenamefont {Ricci}, \citenamefont {Rogers},\ and\ \citenamefont
  {Dorland}(2006)}]{ricci06prl}%
  \BibitemOpen
  \bibfield  {author} {\bibinfo {author} {\bibfnamefont {P.}~\bibnamefont
  {Ricci}}, \bibinfo {author} {\bibfnamefont {B.~N.}\ \bibnamefont {Rogers}}, \
  and\ \bibinfo {author} {\bibfnamefont {W.}~\bibnamefont {Dorland}},\
  }\href@noop {} {\bibfield  {journal} {\bibinfo  {journal} {Phys. Rev. Lett.}\
  }\textbf {\bibinfo {volume} {97}} (\bibinfo {year} {2006})}\BibitemShut
  {NoStop}%
\bibitem [{\citenamefont {Simakov}\ \emph {et~al.}(2001)\citenamefont
  {Simakov}, \citenamefont {Catto}, ,\ and\ \citenamefont
  {Hastie}}]{simakov01}%
  \BibitemOpen
  \bibfield  {author} {\bibinfo {author} {\bibfnamefont {A.~N.}\ \bibnamefont
  {Simakov}}, \bibinfo {author} {\bibfnamefont {P.~J.}\ \bibnamefont {Catto}},
  , \ and\ \bibinfo {author} {\bibfnamefont {R.~J.}\ \bibnamefont {Hastie}},\
  }\href@noop {} {\bibfield  {journal} {\bibinfo  {journal} {Phys. Plasmas}\
  }\textbf {\bibinfo {volume} {8}},\ \bibinfo {pages} {4414} (\bibinfo {year}
  {2001})}\BibitemShut {NoStop}%
\bibitem [{\citenamefont {Angus}, \citenamefont {Dorf},\ and\ \citenamefont
  {Geyko}(2019)}]{angus19}%
  \BibitemOpen
  \bibfield  {author} {\bibinfo {author} {\bibfnamefont {J.~R.}\ \bibnamefont
  {Angus}}, \bibinfo {author} {\bibfnamefont {M.}~\bibnamefont {Dorf}}, \ and\
  \bibinfo {author} {\bibfnamefont {V.~I.}\ \bibnamefont {Geyko}},\ }\href@noop
  {} {\bibfield  {journal} {\bibinfo  {journal} {Phys. Plasmas}\ }\textbf
  {\bibinfo {volume} {26}} (\bibinfo {year} {2019})}\BibitemShut {NoStop}%
\bibitem [{\citenamefont {Shumlak}\ \emph {et~al.}(2001)\citenamefont
  {Shumlak}, \citenamefont {Golingo}, \citenamefont {Nelson},\ and\
  \citenamefont {Hartog}}]{shumlak01}%
  \BibitemOpen
  \bibfield  {author} {\bibinfo {author} {\bibfnamefont {U.}~\bibnamefont
  {Shumlak}}, \bibinfo {author} {\bibfnamefont {R.~P.}\ \bibnamefont
  {Golingo}}, \bibinfo {author} {\bibfnamefont {B.~A.}\ \bibnamefont {Nelson}},
  \ and\ \bibinfo {author} {\bibfnamefont {D.~J.~D.}\ \bibnamefont {Hartog}},\
  }\href@noop {} {\bibfield  {journal} {\bibinfo  {journal} {Phys. Rev. Lett.}\
  }\textbf {\bibinfo {volume} {87}},\ \bibinfo {pages} {205005} (\bibinfo
  {year} {2001})}\BibitemShut {NoStop}%
\bibitem [{\citenamefont {Shumlak}\ \emph {et~al.}(2003)\citenamefont
  {Shumlak}, \citenamefont {Nelson}, \citenamefont {Golingo}, \citenamefont
  {Jackson}, \citenamefont {Crawford},\ and\ \citenamefont
  {Hartog}}]{shumlak03}%
  \BibitemOpen
  \bibfield  {author} {\bibinfo {author} {\bibfnamefont {U.}~\bibnamefont
  {Shumlak}}, \bibinfo {author} {\bibfnamefont {B.~A.}\ \bibnamefont {Nelson}},
  \bibinfo {author} {\bibfnamefont {R.~P.}\ \bibnamefont {Golingo}}, \bibinfo
  {author} {\bibfnamefont {S.~L.}\ \bibnamefont {Jackson}}, \bibinfo {author}
  {\bibfnamefont {E.~A.}\ \bibnamefont {Crawford}}, \ and\ \bibinfo {author}
  {\bibfnamefont {D.~J.~D.}\ \bibnamefont {Hartog}},\ }\href@noop {} {\bibfield
   {journal} {\bibinfo  {journal} {Phys. Plasmas}\ }\textbf {\bibinfo {volume}
  {10}},\ \bibinfo {pages} {1683} (\bibinfo {year} {2003})}\BibitemShut
  {NoStop}%
\bibitem [{\citenamefont {Zhang}\ \emph {et~al.}(2019)\citenamefont {Zhang},
  \citenamefont {Shumlak}, \citenamefont {Nelson}, \citenamefont {Golingo},
  \citenamefont {Stepanov}, \citenamefont {Claveau}, \citenamefont {Forbes},
  \citenamefont {Draper}, \citenamefont {Mitrani}, \citenamefont {McLean},
  \citenamefont {Tummel}, \citenamefont {Higginson},\ and\ \citenamefont
  {Cooper}}]{zhang19}%
  \BibitemOpen
  \bibfield  {author} {\bibinfo {author} {\bibfnamefont {Y.}~\bibnamefont
  {Zhang}}, \bibinfo {author} {\bibfnamefont {U.}~\bibnamefont {Shumlak}},
  \bibinfo {author} {\bibfnamefont {B.~A.}\ \bibnamefont {Nelson}}, \bibinfo
  {author} {\bibfnamefont {R.~P.}\ \bibnamefont {Golingo}}, \bibinfo {author}
  {\bibfnamefont {T.~R. W. A.~D.}\ \bibnamefont {Stepanov}}, \bibinfo {author}
  {\bibfnamefont {E.~L.}\ \bibnamefont {Claveau}}, \bibinfo {author}
  {\bibfnamefont {E.~G.}\ \bibnamefont {Forbes}}, \bibinfo {author}
  {\bibfnamefont {Z.~T.}\ \bibnamefont {Draper}}, \bibinfo {author}
  {\bibfnamefont {J.~M.}\ \bibnamefont {Mitrani}}, \bibinfo {author}
  {\bibfnamefont {H.~S.}\ \bibnamefont {McLean}}, \bibinfo {author}
  {\bibfnamefont {K.~K.}\ \bibnamefont {Tummel}}, \bibinfo {author}
  {\bibfnamefont {D.~P.}\ \bibnamefont {Higginson}}, \ and\ \bibinfo {author}
  {\bibfnamefont {C.~M.}\ \bibnamefont {Cooper}},\ }\href@noop {} {\bibfield
  {journal} {\bibinfo  {journal} {Phys. Rev. Lett.}\ }\textbf {\bibinfo
  {volume} {122}} (\bibinfo {year} {2019})}\BibitemShut {NoStop}%
\bibitem [{\citenamefont {Shumlak}(2020)}]{shumlak20}%
  \BibitemOpen
  \bibfield  {author} {\bibinfo {author} {\bibfnamefont {U.}~\bibnamefont
  {Shumlak}},\ }\href@noop {} {\bibfield  {journal} {\bibinfo  {journal} {J.
  Appl. Phys.}\ }\textbf {\bibinfo {volume} {127}} (\bibinfo {year}
  {2020})}\BibitemShut {NoStop}%
\bibitem [{\citenamefont {Golingo}, \citenamefont {Shumlak},\ and\
  \citenamefont {Nelson}(2005)}]{golingo05}%
  \BibitemOpen
  \bibfield  {author} {\bibinfo {author} {\bibfnamefont {R.~P.}\ \bibnamefont
  {Golingo}}, \bibinfo {author} {\bibfnamefont {U.}~\bibnamefont {Shumlak}}, \
  and\ \bibinfo {author} {\bibfnamefont {B.~A.}\ \bibnamefont {Nelson}},\
  }\href@noop {} {\bibfield  {journal} {\bibinfo  {journal} {Phys. Plasmas}\
  }\textbf {\bibinfo {volume} {12}} (\bibinfo {year} {2005})}\BibitemShut
  {NoStop}%
\bibitem [{\citenamefont {Shumlak}\ and\ \citenamefont
  {Hartman}(1995)}]{shumlak95}%
  \BibitemOpen
  \bibfield  {author} {\bibinfo {author} {\bibfnamefont {U.}~\bibnamefont
  {Shumlak}}\ and\ \bibinfo {author} {\bibfnamefont {C.~W.}\ \bibnamefont
  {Hartman}},\ }\href@noop {} {\bibfield  {journal} {\bibinfo  {journal} {Phys.
  Rev. Lett.}\ }\textbf {\bibinfo {volume} {75}},\ \bibinfo {pages} {3285}
  (\bibinfo {year} {1995})}\BibitemShut {NoStop}%
\bibitem [{\citenamefont {Arber}\ and\ \citenamefont {Howell}(1996)}]{arber96}%
  \BibitemOpen
  \bibfield  {author} {\bibinfo {author} {\bibfnamefont {T.~D.}\ \bibnamefont
  {Arber}}\ and\ \bibinfo {author} {\bibfnamefont {D.~F.}\ \bibnamefont
  {Howell}},\ }\href@noop {} {\bibfield  {journal} {\bibinfo  {journal} {Phys.
  Plasmas}\ }\textbf {\bibinfo {volume} {3}} (\bibinfo {year}
  {1996})}\BibitemShut {NoStop}%
\bibitem [{\citenamefont {Angus}, \citenamefont {Dorf},\ and\ \citenamefont
  {Geyko}(2020)}]{angus20dg}%
  \BibitemOpen
  \bibfield  {author} {\bibinfo {author} {\bibfnamefont {J.~R.}\ \bibnamefont
  {Angus}}, \bibinfo {author} {\bibfnamefont {M.~A.}\ \bibnamefont {Dorf}}, \
  and\ \bibinfo {author} {\bibfnamefont {V.~I.}\ \bibnamefont {Geyko}},\
  }\href@noop {} {\bibfield  {journal} {\bibinfo  {journal} {Submitted to Phys.
  Plasmas}\ } (\bibinfo {year} {2020})}\BibitemShut {NoStop}%
\bibitem [{\citenamefont {Paraschiv}\ \emph {et~al.}(2010)\citenamefont
  {Paraschiv}, \citenamefont {Bauer}, \citenamefont {Lindemuth},\ and\
  \citenamefont {Makhin}}]{paraschiv10}%
  \BibitemOpen
  \bibfield  {author} {\bibinfo {author} {\bibfnamefont {I.}~\bibnamefont
  {Paraschiv}}, \bibinfo {author} {\bibfnamefont {B.~S.}\ \bibnamefont
  {Bauer}}, \bibinfo {author} {\bibfnamefont {I.~R.}\ \bibnamefont
  {Lindemuth}}, \ and\ \bibinfo {author} {\bibfnamefont {V.}~\bibnamefont
  {Makhin}},\ }\href@noop {} {\bibfield  {journal} {\bibinfo  {journal} {Phys.
  Plasmas}\ }\textbf {\bibinfo {volume} {17}} (\bibinfo {year}
  {2010})}\BibitemShut {NoStop}%
\bibitem [{\citenamefont {Geyko}, \citenamefont {Dorf},\ and\ \citenamefont
  {Angus}(2019)}]{geyko19}%
  \BibitemOpen
  \bibfield  {author} {\bibinfo {author} {\bibfnamefont {V.~I.}\ \bibnamefont
  {Geyko}}, \bibinfo {author} {\bibfnamefont {M.}~\bibnamefont {Dorf}}, \ and\
  \bibinfo {author} {\bibfnamefont {J.~R.}\ \bibnamefont {Angus}},\ }\href@noop
  {} {\bibfield  {journal} {\bibinfo  {journal} {Phys. Plasmas}\ }\textbf
  {\bibinfo {volume} {26}} (\bibinfo {year} {2019})}\BibitemShut {NoStop}%
\bibitem [{\citenamefont {Tummel}\ \emph {et~al.}(2019)\citenamefont {Tummel},
  \citenamefont {Higginson}, \citenamefont {Link}, \citenamefont {Schmidt},
  \citenamefont {McLean}, \citenamefont {Offermann}, \citenamefont {Welch},
  \citenamefont {Clark}, \citenamefont {Shumlak}, \citenamefont {Nelson},\ and\
  \citenamefont {Golingo}}]{tummel19}%
  \BibitemOpen
  \bibfield  {author} {\bibinfo {author} {\bibfnamefont {K.}~\bibnamefont
  {Tummel}}, \bibinfo {author} {\bibfnamefont {D.~P.}\ \bibnamefont
  {Higginson}}, \bibinfo {author} {\bibfnamefont {A.~J.}\ \bibnamefont {Link}},
  \bibinfo {author} {\bibfnamefont {A.~E.~W.}\ \bibnamefont {Schmidt}},
  \bibinfo {author} {\bibfnamefont {H.~S.}\ \bibnamefont {McLean}}, \bibinfo
  {author} {\bibfnamefont {D.~T.}\ \bibnamefont {Offermann}}, \bibinfo {author}
  {\bibfnamefont {D.~R.}\ \bibnamefont {Welch}}, \bibinfo {author}
  {\bibfnamefont {R.~E.}\ \bibnamefont {Clark}}, \bibinfo {author}
  {\bibfnamefont {U.}~\bibnamefont {Shumlak}}, \bibinfo {author} {\bibfnamefont
  {B.~A.}\ \bibnamefont {Nelson}}, \ and\ \bibinfo {author} {\bibfnamefont
  {R.~P.}\ \bibnamefont {Golingo}},\ }\href@noop {} {\bibfield  {journal}
  {\bibinfo  {journal} {Phys. Plasmas}\ }\textbf {\bibinfo {volume} {26}}
  (\bibinfo {year} {2019})}\BibitemShut {NoStop}%
\bibitem [{\citenamefont {Arber}, \citenamefont {Coppins},\ and\ \citenamefont
  {Scheffel}(1994)}]{arber94}%
  \BibitemOpen
  \bibfield  {author} {\bibinfo {author} {\bibfnamefont {T.~D.}\ \bibnamefont
  {Arber}}, \bibinfo {author} {\bibfnamefont {M.}~\bibnamefont {Coppins}}, \
  and\ \bibinfo {author} {\bibfnamefont {J.}~\bibnamefont {Scheffel}},\
  }\href@noop {} {\bibfield  {journal} {\bibinfo  {journal} {Phys. Rev. Lett.}\
  }\textbf {\bibinfo {volume} {72}} (\bibinfo {year} {1994})}\BibitemShut
  {NoStop}%
\bibitem [{\citenamefont {Bennett}(1934)}]{bennett34}%
  \BibitemOpen
  \bibfield  {author} {\bibinfo {author} {\bibfnamefont {W.~H.}\ \bibnamefont
  {Bennett}},\ }\href@noop {} {\bibfield  {journal} {\bibinfo  {journal} {Phys.
  Rev.}\ }\textbf {\bibinfo {volume} {45}},\ \bibinfo {pages} {890} (\bibinfo
  {year} {1934})}\BibitemShut {NoStop}%
\bibitem [{\citenamefont {Bennett}(1955)}]{bennett55}%
  \BibitemOpen
  \bibfield  {author} {\bibinfo {author} {\bibfnamefont {W.~H.}\ \bibnamefont
  {Bennett}},\ }\href@noop {} {\bibfield  {journal} {\bibinfo  {journal} {Phys.
  Rev.}\ }\textbf {\bibinfo {volume} {98}},\ \bibinfo {pages} {1584} (\bibinfo
  {year} {1955})}\BibitemShut {NoStop}%
\bibitem [{\citenamefont {Dorf}\ \emph {et~al.}(2013)\citenamefont {Dorf},
  \citenamefont {Cohen}, \citenamefont {Dorr}, \citenamefont {Rognlien},
  \citenamefont {Hittinger}, \citenamefont {Compton}, \citenamefont {Colella},
  \citenamefont {Martin},\ and\ \citenamefont {McCorquodale}}]{dorf13}%
  \BibitemOpen
  \bibfield  {author} {\bibinfo {author} {\bibfnamefont {M.~A.}\ \bibnamefont
  {Dorf}}, \bibinfo {author} {\bibfnamefont {R.~H.}\ \bibnamefont {Cohen}},
  \bibinfo {author} {\bibfnamefont {M.}~\bibnamefont {Dorr}}, \bibinfo {author}
  {\bibfnamefont {T.}~\bibnamefont {Rognlien}}, \bibinfo {author}
  {\bibfnamefont {J.}~\bibnamefont {Hittinger}}, \bibinfo {author}
  {\bibfnamefont {J.}~\bibnamefont {Compton}}, \bibinfo {author} {\bibfnamefont
  {P.}~\bibnamefont {Colella}}, \bibinfo {author} {\bibfnamefont
  {D.}~\bibnamefont {Martin}}, \ and\ \bibinfo {author} {\bibfnamefont
  {P.}~\bibnamefont {McCorquodale}},\ }\href@noop {} {\bibfield  {journal}
  {\bibinfo  {journal} {Phys. Plasmas}\ }\textbf {\bibinfo {volume} {20}}
  (\bibinfo {year} {2013})}\BibitemShut {NoStop}%
\bibitem [{\citenamefont {Chew}, \citenamefont {Goldberger},\ and\
  \citenamefont {Low}(1956)}]{chew56}%
  \BibitemOpen
  \bibfield  {author} {\bibinfo {author} {\bibfnamefont {G.~F.}\ \bibnamefont
  {Chew}}, \bibinfo {author} {\bibfnamefont {M.~L.}\ \bibnamefont
  {Goldberger}}, \ and\ \bibinfo {author} {\bibfnamefont {F.~E.}\ \bibnamefont
  {Low}},\ }\href@noop {} {\bibfield  {journal} {\bibinfo  {journal} {Proc.
  Roy. Soc.}\ }\textbf {\bibinfo {volume} {A236}},\ \bibinfo {pages} {112}
  (\bibinfo {year} {1956})}\BibitemShut {NoStop}%
\bibitem [{\citenamefont {Bellan}(2006)}]{bellan06}%
  \BibitemOpen
  \bibfield  {author} {\bibinfo {author} {\bibfnamefont {P.}~\bibnamefont
  {Bellan}},\ }\href@noop {} {\emph {\bibinfo {title} {Fundamentals of Plasma
  Physics}}}\ (\bibinfo  {publisher} {Cambridge University Press},\ \bibinfo
  {year} {2006})\BibitemShut {NoStop}%
\end{thebibliography}%

\end{document}